# Stable Dynamic Predictive Clustering (SDPC) Protocol for Vehicular Ad hoc Network

Mohammad Mukhtaruzzaman and Mohammed Atiquzzaman

*School of Computer Science*

*University of Oklahoma*

Norman, OK-73019

mukhtar@ou.edu, atiq@ou.edu

*Abstract*— Vehicular communication is an essential part of a smart city. Scalability is a major issue for vehicular communication, specially, when the number of vehicles increases at any given point. Vehicles also suffer some other problems such as broadcast problem. Clustering can solve the issues of vehicular ad hoc network (VANET); however, due to the high mobility of the vehicles, clustering in VANET suffers stability issue. Previously proposed clustering algorithms for VANET are optimized for either straight road or for intersection. Moreover, the absence of the intelligent use of a combination of the mobility parameters, such as direction, movement, position, velocity, degree of vehicle, movement at the intersection etc., results in cluster stability issues. A dynamic clustering algorithm considering the efficient use of all the mobility parameters can solve the stability problem in VANET. To achieve higher stability for VANET, a novel robust and dynamic clustering algorithm stable dynamic predictive clustering (SDPC) for VANET is proposed in this paper. In contrast to previous studies, vehicle relative velocity, vehicle position, vehicle distance, transmission range, and vehicle density are considered in the creation of a cluster, whereas relative distance, movement at the intersection, degree of vehicles are considered to select the cluster head. From the mobility parameters the future road scenario is constructed. The cluster is created, and the cluster head is selected based on the future construction of the road. The performance of SDPC is compared in various road topologies with four existing VANET clustering algorithms in terms of the average cluster head change rate, the average cluster head duration, the average cluster member duration, and the ratio of clustering overhead in terms of total packet transmission. The simulation result shows SDPC outperforms the existing algorithms and achieved better clustering stability.

## 1 Introduction

Vehicular ad hoc networks (VANETs) are used to improve road safety. In VANET, vehicles transmit road information to other vehicles and receive road information from other vehicles using vehicle-to-vehicle (V2V) communication. Vehicles communicate with different infrastructures, for example road-side units (RSUs), using vehicle-to-infrastructure (V2I) communication. V2V and V2I communication combinedly termed as vehicle-to-everything (V2X) communication.

Number of vehicles on the road is increasing day by day. When the number of vehicles increases, VANETs suffer broadcast storm problem and scalability issue. Intersections are very common in today's road structure where scalability is more important. Vehicles are fast running object and for road safety, faster vehicle-to-vehicle communication is required. In the absence of any central infrastructure, a hierarchy of vehicles was needed. Clustering is proposed for VANET to solve these problems.

In VANET clustering, vehicles are divided into several groups based on the similar features of the vehicles such as velocity, position, distance, degree of vehicles etc.

Velocity is one of the key parameters in VANET clustering. In this case, velocity of the vehicles among the clusters are measured and relative velocity is calculated. The vehicle with the lowest relative velocity is selected as the cluster head (CH). Using the transmission range (TR) of the CH, the cluster is created.

Position of a vehicle among a group of vehicles is also an important parameter to form cluster. Position of the vehicles among the clusters are measured and relatively stable vehicle is selected as the CH. Using the TR of the CH, the cluster is created.

Degree of vehicle is also used to create cluster. The number of vehicles within the TR of a vehicle is called the degree of a vehicle. The degree of vehicles is counted and the vehicle with the highest degree of vehicle is selected as the CH. Using the TR of the CH, the cluster is created.

The vehicles are highly moving object and change their direction frequently while TR is limited; as a result, clusters break down frequently.

In the literature, many clustering algorithms have been proposed to minimize break down of the clusters. Some algorithms [] select the vehicle with the lowest relative velocity as the CH so that the member vehicles need more time to go out of the TR of the CH. After selecting the CH, a cluster is formed including the vehicles within the TR of the CH.

Some algorithms [1,2] use position of the vehicles as the basis of clustering. Position of the vehicles among the clusters are measured and relatively center vehicle is selected as the CH so that the member vehicles need to cross more distance to go out of the cluster. After selecting the CH, a cluster is formed including the vehicles within the TR of the CH.

Some algorithms [3] use degree of vehicles as the basis of clustering. The degree of vehicles is counted and the vehicle with the highest degree of vehicle is selected as the CH so that a greater number of vehicles can be accommodated within the cluster to reduce the number of clusters. After selecting the CH, a cluster is formed including the vehicles within the TR of the CH.

The existing algorithms are kind of effective in straight road without any intersection; however, intersections are not given proper attention in the existing algorithms. As a result, clusters of the existing algorithms break down at the intersection of the roads.

Clustering was proposed to solve the scalability issue and scalability becomes more significant at the intersection where the existing algorithms do not work properly at the intersection. Therefore, we need new clustering algorithm for VANET which will consider the intersections of the road during CH selection and cluster formation process to tackle the scalability issue in VANET. The scalability is measured with the average number of clusters on the road at any given time.

We see the existing algorithms did not consider the intersections, as a result, the algorithms will not work at the intersection. For example, five vehicles are running at a road. We can create cluster here. Even without creating cluster for five vehicles, we can set up effective communication among five vehicles. Moreover, creating cluster among five vehicles require lesser clustering overhead. Even frequent cluster breaking will not cause any effect on other vehicles the roads that are not in the TR of these vehicles.

Now, many vehicles come within the TR at the intersection and many vehicles run out of the TR. Instead of five vehicles, now twenty vehicles are within the TR of the vehicles. Therefore, any vehicle is now transmitting and receiving the message with twenty vehicles. To create cluster at this point will create more clustering overhead. Besides, any change in the cluster will cause a huge number of message transmission. Consequently, scalability problem will be severe.

To overcome the scalability issue at the intersection, we need to consider the movement of the vehicles at the intersection during the CH selection and formation of cluster so that a lesser number of vehicles joining and leaving happens at the intersection.

We have utilized the mobility parameters, considered intersection, and constructed a future position of the vehicles to make cluster efficient and stable. We consider velocity, acceleration, movement of the vehicle, movement of the vehicle after next intersection, degree of vehicle etc. We also consider the degree of vehicles of similarly suitable candidates. As a result, our algorithm can create more efficient cluster.

We have achieved better results for the average CH change rate, the average CH duration, the average CM duration, and the clustering overhead which reflects that SDFC can provide better clustering stability regardless of road topology.

Three states of the cars are used: Entry state (EN), Cluster Head (CH), and Cluster Member (CM). CH is the head of a cluster, CM is a member of a cluster, and EN is the initial state. A vehicle can access its neighbor vehicles within its transmission range (TR). The algorithm uses one-hop clustering as shown in Fig. 1. Hence. clustering is limited by the TR of the CH of a cluster.

The rest of the paper is organized as follows. Section 2 describes the related works and SDPC is presented in Section 3. Section 4 describes the performance metrics used to evaluate. Section 5 discusses the simulation results and compares the performance. Finally, concluding remarks are presented in Section 6.

## 2 RELATED WORKS

Many clustering algorithms have been proposed in the literature as shown in Table I. Some algorithms are intelligence-based, and some are mobility-based. Intelligence-based clustering algorithms are two types: machine learning-based algorithms and fuzzy logic-based algorithm.

K-means algorithm is proposed by Taherkhani et al. [9] for VANET clustering considering the distance and direction of vehicles with some other features to create cluster and to select the CH; however, k-means algorithm requires the number of clusters as the initial input that is not suitable for the dynamicity of VANET. A modified K-Means algorithm, optimized for highway environment, is used by Kandali et al. [10] where random selectin of the number of clusters and the initial CH is prevented. Vehicles are assigned into clusters based on link reliability depending on vehicle density, relative speed, and distance. The CH is selected in this routing protocol based on velocity, degree of node, and the buffer size; however, no major cluster stability metrics are evaluated. A hierarchical algorithm is proposed by Bhosale et al. [11] to overcome the limitations of the k-means algorithm where the direction and speed of the vehicles are given as the input in the clustering process. The algorithm did not present an evaluation of the stability metrics. A nature-inspired optimization algorithm [12] is proposed where only the number of clusters is evaluated for multiple scenarios what is relatively a minor metric to evaluate the stability of an algorithm. A combination of fuzzy logic and Q-learning [13] is used where the slowest vehicle is selected as the CH to avoid frequent change of CH without evaluating any stability metrics. Sharma et al. [14] used the Dolphin swarm algorithm with fuzzy logic creating multiple CHs at a time to reduce the overload from a single CH; however, no stability metrics are evaluated.

Some mobility-based algorithms are proposed for VANET clustering. Qi et al. [4] proposed a clustering algorithm to improve the cluster stability based on the social patterns. Based on the historical movement pattern vehicles are grouped in the same cluster who follow the same route. Relative speed and inter-vehicle distance are considered to select the CH, the strategy is evaluated based on cluster lifetime and clustering overhead only. The lifetime of a cluster is important but cannot be the only measurement to evaluate cluster stability. If CM loses connection with the CH frequently, clusters will be unstable. Therefore, CM-related metrics need to be evaluated to measure cluster stability. The most reliable vehicle is selected as the CH in [5] based on the relative velocity of the vehicles. Position, velocity, and direction are considered to form a cluster; however, the stability metrics are not evaluated. Bersali et al. [7] proposed a new collaborative clustering algorithm based on node score where a vehicle with high node score is selected as the CH. The node score is calculated using degree of node, distance, link stability, average relative speed, average relative acceleration etc. They also introduced a vice CH for better cluster stability; however, how a vice CH improves cluster stability is not described properly. Moreover, only three metrics have been evaluated that is not sufficient to prove the performance of the algorithm.

VMaSC was proposed by Ucar et al. [6] where a cluster is formed based on the relative velocity. The relative velocity is measured among the neighbors of the N-hop to create the clusters and the vehicle with the lowest velocity is selected as the CH. A new member is added to the neighbor CM in a multi-hop manner. Merging is allowed in this scheme if two CHs spend a certain amount of time within their range. The CH acts as a dual-interface node. The CH communicates with CMs via V2V and connects the cluster to the cellular network via V2I communication; however, the results are specific to a particular set of parameters under which the simulation is carried out. Also, CM-related metrics are not optimized. Ren et al. [8] considered moving direction, relative vehicle position, and link lifetime to create clusters. Temporary cluster head and a safe distance threshold introduced to increase cluster stability. The vehicle which is in the center of a cluster is selected as the CH so that a CM spend more travel time to leave a cluster. When two clusters come to a close distance, they can merge. Most of the stability metrics have been evaluated; however, in contrast to [6], this scheme is optimized for CM metrics. Hence, CH-related metrics such as CH duration, number of CH change, etc., are not optimized. Moreover, many times it creates a large number of single clusters.

JCV [15] used various parameters to create cluster and to select the CH. The average relative velocity, the degree of vehicles, the movement of the vehicle at the intersection etc are considered and they evaluated many metrics that shows their superior performance; however, they did not compare with any other intersection-based clustering algorithm, even though they presented comparison with two algorithms.

Table 1. Notations used

| NOTATION | DESCRIPTION |
|---|---|
| CH | CLUSTER HEAD |
| CM | CLUSTER MEMBER |
| TM | TEMPORAL STATE |
| TR | TRANSMISSION RANGE |
| CH_TIMER | CLUSTER HEAD STATE TIMER |
| CM_TIMER | CLUSTER MEMBER STATE TIMER |
| EN_TIMER | ENTRY STATE TIMER |
| VEH_INFO | VEHICLE INFO |
| EN_REQ | MESSAGE FROM EN |
| CH_RESP | RESPONSE FROM A CH |
| VEH_ADV | PERIODIC VEHICLE ADVERTISE MESSAGE |
| REL_DIST | RELATIVE DISTANCE |
| AVG_REL_DIST | AVERAGE RELATIVE DISTANCE |
| VEH_RANK | RANKING OF A VEHICLE |
| STATUS | STATUS OF A VEHICLE |

# 3  PROPOSED CLUSTERING ALGORITHM

In the proposed algorithm, vehicles can remain in one of three states: Temporal (TM), Cluster Head (CH), and Cluster Member (CM). TM is a temporary state when vehicle enters the road, collect information about other vehicles within the transmission range, and tries to join in any existing cluster. CH is the head of a cluster and CMs are the member of the cluster. Vehicles communicate its neighbor vehicles within its transmission range using on-board unit. All the vehicles are equipped with global positioning system (GPS).

## 3.1   State of Vehicles

When a vehicle enters the road, it enters with TM state. It is a temporary state when the vehicle collects information from other vehicles withing its transmission range. If any vehicle is available within the transmission range, the vehicle tries to form a cluster.

If a vehicle is the most suitable candidate to be selected as the CH, then the vehicle changes its status to CH and advertise to all the vehicle in its transmission range. If there is no vehicle in its transmission range, then the vehicle remains as a single cluster, i.e., a CH without any member.

If a vehicle finds any vehicle in its transmission range which is more suitable to be CH, then the vehicle requests the CH to join the cluster. After receiving acknowledge message from CH, the vehicle changes its status to CM.

### 3.2 State Transitions

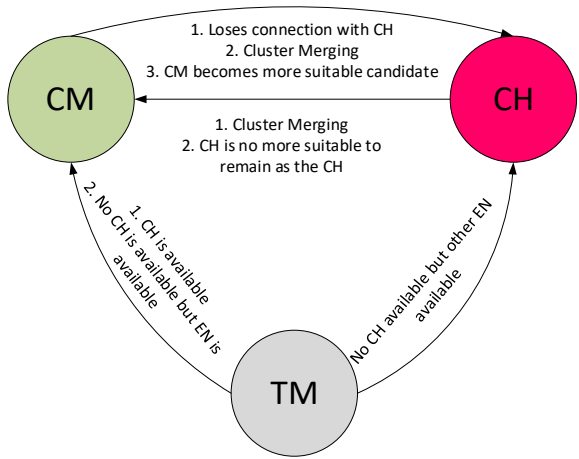

Fig. 1. State transition diagram.

Vehicles enter as TM, sets a timer, and broadcasts VEH_ADV message periodically to inform its presence on the road. Fig. 1 illustrates the state transitions of the proposed algorithm.

From TM state, a vehicle can convert to a CH or CM. If there is no CH available but at least one TM is available within the TR of the TM and the TM is a better candidate for CH, then a cluster will be created, and the TM will be the CH of the new cluster. The vehicle enters from TM to CM state if it receives an acknowledgment message from a CH of an existing cluster. If there is no CH available but at least one TM is available within the TR of the TM, and the other TM is a better candidate to become a CH, then the TM will be a CM of the new cluster.

A CH can transform to CM if condition satisfies. If two CHs come within the TR of each other and decide to merge into one cluster, then one or both CH changes their STATUS into CM. The same is applied if an TM comes within the TR. Secondly, due to the dynamicity of the vehicles, if a CM becomes a better candidate to be the CH, the present CH downgrades its STATUS to CM.

A CM can transform to CH satisfying certain criteria. A CM can become a CH if the CM loses connection with its CH and some vehicles are present within TR and this CM is the best candidate to become the CH. Secondly, during the cluster merging process, a CM can be elected as the CH of the new cluster if the CM is the best candidate compared to existing CHs and the rest of the vehicles. Thirdly, a CM can become a CH of its cluster if it becomes the best candidate than the rest of the vehicles due to vehicles' movement.

### 3.3 Cluster Formation

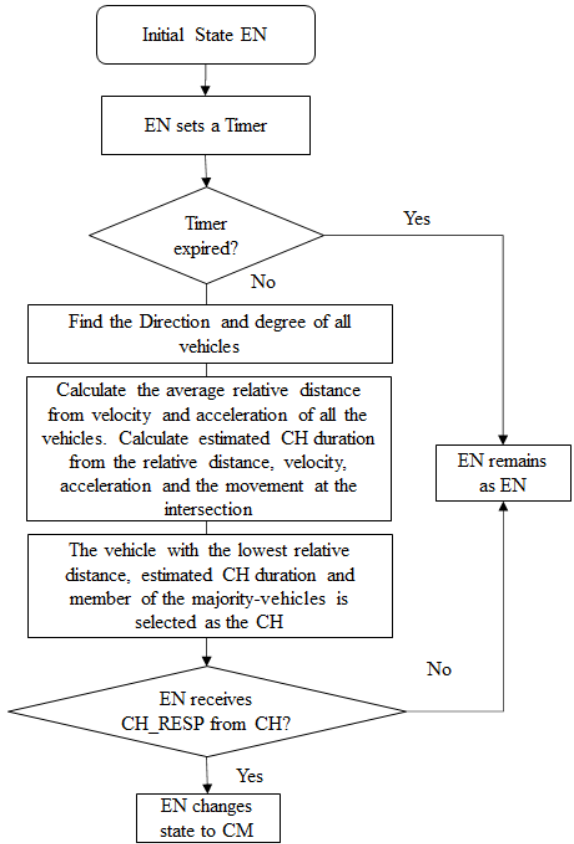

Fig. 2. Cluster formation process.

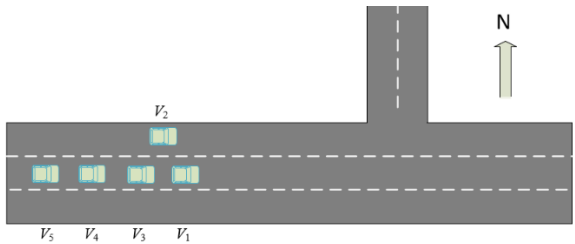

Fig. 3. Illustration of movement of the vehicles after the next intersection.

At the next intersection, $V_1$, $V_3$, $V_4$, $V_5$ will run towards the east direction while $V_2$ will be running into the north direction. The number of vehicles in the east direction is four which is greater than the number of vehicles in the north direction which is one. Therefore, the CH will be selected from the vehicles who will be running into the east direction. $V_2$ is a minority vehicle will not be considered for the candidacy of CH; however, it can join the cluster as a CM.

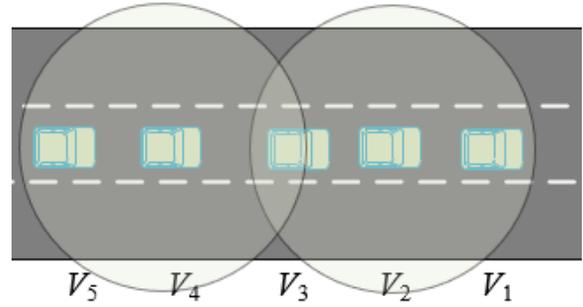

Fig. 4. Degree of node consideration.

The degree of vehicle is the number of vehicles that can be reached directly in one-hop distance. The degree of vehicle is applied when more than one vehicle becomes equally qualified candidates to be the CH and the calculation of the degree of vehicle is different. The degree of node will be calculated for those vehicles who has the coverage of the first vehicle of the probable cluster. In this way, the number of stranded vehicles, i.e., the number of single clusters can be reduced. In Fig.4, degree of vehicle of $V_3$ is 2, but if $V_3$ is selected as the CH, then $V_1$ and $V_5$ will not have any vehicle to form a cluster. On the other hand, if $V_2$ is selected as the CH who has the same degree of node as $V_3$, but $V_4$ and $V_5$ will have the opportunity to form a new cluster, hence, a greater number of vehicles will be in clustered state. All the vehicles, including any vehicle who have a different moving direction after the next intersection, will be considered to count the degree of vehicle of a vehicle; however, from the previous paragraph, a vehicle with different moving direction after the next intersection will not be a candidate to be a CH. The minority vehicle will be considered to become the CM of the new cluster.

### 3.4 CH Selection

CH selection and cluster formation will be performed concurrently. The most stable vehicles will be determined and will be selected as the CH and cluster will be formed starting from the front end.

In this algorithm, geographical front vehicle from a group of vehicles where clustering algorithm will be applied is termed as the first vehicle. The following notations have been used.

$\bar{u}_i(t)$ = Average Relative velocity of $V_i$ with respect to other vehicles at time $t = \frac{\sum_{j=1}^{n}(u_{i,j}(t))}{n-1}$ where $i \neq j$.

Velocity, acceleration, and distance are considered to calculate $\bar{s}_{ij}(t)$ for all the vehicles. The vehicle with $\min(\bar{s}_i(t))$ will be selected as the CH of the cluster.

For all the vehicles, $s_{ij}(t)$ will be calculated, at any time, $t$,

$$s_{ij}(t) = s_{ij}(0) + \left(u_i(0) - u_j(0)\right)t + \int_0^t \int_0^t \left(a_i(t) - a_j(t)\right) dt\, dt$$

Therefore, the minimum of $\bar{s}_{ij}(t)$ is, $\min(\bar{s}_i(t))$, where $(1 \leq i \leq n)$

If multiple candidates are found, then the degree of vehicle will be applied, i.e., the vehicle with $\min(\bar{s}_i(t))$ and $\max(degree\ of\ vehicle)$ will be selected as the CH. Each vehicle advertises its expected CH lifetime during cluster formation process. If we have multiple candidates, then the vehicle with higher probable CH lifetime will be selected as the CH.

CH selection and Cluster formation Algorithm:

For any group of vehicles DO

1. Find the degree of vehicle for all vehicles.
2. Find the direction for all vehicles at the next intersection.
3. Calculate $\bar{s}_i(t)$ for all vehicle (where $t = 0$).
4. CH Selection: Select the CH if the vehicle satisfies the following conditions at $t = 0$:
5. Create a cluster based on the *TR* of the CH selected at STEP 4.
6. Select a new first vehicle from the rest of the vehicles excluding the vehicles who are already in a cluster.
7. Repeat STEP 4 to 6.

### 3.5 Gateway Selection

Cluster gateways are two CMs who maintains communication with other clusters. Two vehicles from the front and the back will be selected as the CGs. After creating the cluster, CH will appoint two CMs as the CGs.

### 3.6 Cluster Merging

If two clusters overlap each other for a predefined time period, two CH will initiate cluster merging process. Two CHs will transmit information of all the member vehicles. If a combined cluster is possible, two CH will start the merging process. Both the CH will transmit information of the CMs to the newly selected CH. Upon receiving the request from the CHs, the new CH will send acknowledgement messages to the old CHs. The old CHs will send the response messages to the new CH as well as to all the CMs about its status as the CH.

### 3.7 Cluster Leaving

When a CM wants to leave a cluster, the CM will send a request to the CH. Upon receiving acknowledgement message from the CH, the CM will leave a cluster. Besides, if the CH does not receive any regular message from any CM for a specified time, the vehicle is considered to be left the cluster. The CH will delete the entry for the CM and advertise to the clusters.

## 4 SIMULAITON

### *4.1 Assumptions*

**Topology**

The road topology used for simulation is shown in Fig. 5. Roads contain multiple intersections. Total road length is 60 km; however, no vehicle runs for entire 60 km. The lower four nodes are the starting points for the vehicles and the upper four nodes are the end points. Vehicles start from one of the four starting points and ends at one of the four end points. At the intersection, some vehicles run straight, some vehicles run to the left direction and some vehicles run to the right direction, i.e., vehicles running from the west to the east direction can run towards the east, north, or east direction at the intersection. No vehicle runs at the west direction. A vehicle running from south to north direction can run towards the north or east direction at the intersection. Similarly, a vehicle running from the north to the south direction can run towards the south or east direction at the intersection.

**Number of Lanes**

The road used in the simulation consists of two-lanes two-way roads.

**Overtaking**

Vehicles can overtake other vehicles. The overtaking decision of the vehicles is calculated using the distance, velocity of the two vehicles, and acceleration-deceleration of the vehicles.

**Entry speed**

Entry speed varies from 10 m/s to 30 m/s depending on the maximum velocity. Entry speed does not exceed the maximum velocity.

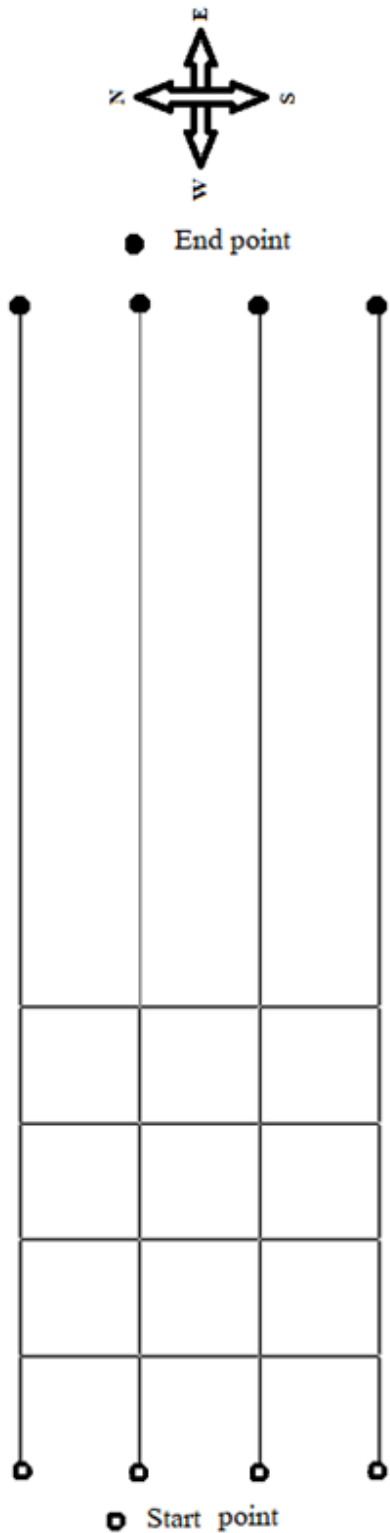

Fig. 5. Road topology used in simulation.

Mohammad Mukhtaruzzaman

**Maximum velocity**

Vehicles run at different maximum velocities from 10 m/s to 35 m/s. Some vehicles achieve exact maximum speed while some vehicles achieve some random maximum speed.

**Mobility model**

Car-following model has been used where vehicles maintain a gap two-seconds gap between two vehicles.

**Direction**

Vehicle can change their direction at the intersection. If the direction is changed, vehicles slow down at the intersection at a random speed and accelerate to the maximum speed after the intersection.

**Entry process**

Vehicles enter the road following a Poisson process at the rate of 2 vehicles/second using the car-following model.

**Result calculation**

After all the vehicles have entered the road, the simulation results are calculated. Total simulation time is 300 seconds. 100 vehicles are divided into three groups: 34, 33, and 33 in each group. The first group of vehicles run from the east to the west direction only. The second group of vehicles occasionally change the direction, and the third group of vehicles frequently change their direction at the intersection. The first group of vehicles enter the road at the rate of 2 vehicles/seconds and after all vehicles reaching at the maximum speed, the results are calculated from 51 seconds to 250 seconds, i.e., 200 seconds. These results are converted into 100 seconds. Then the second group of vehicles enter the road, and the results are calculated for 200 seconds, and the results are converted into 100 seconds. Similarly, the third group of vehicles enter the road, results are calculated for 200 seconds, and converted into 100 seconds. Finally, the results from each group are used to construct the final results for 300 seconds.

TABLE I

SIMULATION PARAMETERS

| NOTATION | DESCRIPTION |
|---|---|
| SIMULATION TIME | 300S |
| MAXIMUM VELOCITY | 10 – 35 M/S |
| NUMBER OF VEHICLES | 100 |
| TRAMISSION RANGE | 200 M |
| ACCELERATION RATE | 10 M/S$^2$ |
| DECELARATION RATE | 5 M/S$^2$ |
| TRAFFIC FLOW RATE | 2 VEHICLES/SECOND |
| MAC PROTOCOL | IEEE 802.11P |
| *VEH_ADV* PERIOD | 200 MS |
| *VEH_ADV* SIZE | 64 BYTES |
| *EN_TIMER* | 2 S |
| *CM_TIMER* | 2 S |
| *CH_TIMER* | 2 S |
| *MERGE_TIMER* | 2 S |
| MOBILITY MODEL | CAR-FOLLOWING MODEL |

*4.2 Simulation Setup*

The performance of SDPC has been compared with four other VANET clustering algorithms: VMaSC, DCV, ECHS and IBC. For traffic generation, we used Simulation of Urban Mobility (SUMO) [16] and for clustering, we used the cloud based VANET simulator (CVANETSIM) [17].

*4.3 Performance Metrics*

The performance metrics used to evaluate MM algorithm are described below.

Mohammad Mukhtaruzzaman

CH change rate per second: This is the number of state transitions from CH to another state per unit time. It reflects the longevity of a cluster. A lower CH change rate is preferable for the stability of VANET clustering. For cluster stability, this is the most important metric. New cluster formation requires a lot of message transmission among the vehicles. The main purpose of clustering in VANET is reflected by this metric.

Average CH duration: This is the time between a vehicle becoming a CH and subsequently changing to another state. A longer average CH duration is highly expected.

Average CM duration: This is the average time a vehicle spends as a CM. Average CM duration does not have significant effect on cluster stability.

Clustering overhead: This is the ratio of the number of clustering related packet to the total number of packets. This is useful to know how many extra packets are generated by the vehicles due to clustering compared to the total number of packet transmission.

## 5 RESULTS AND ANALYSIS

The comparison results from simulation of SDPC, DCV, VMaSC, IBC, and ECHS are described below.

### 5.1 CH Change Rate(per second)

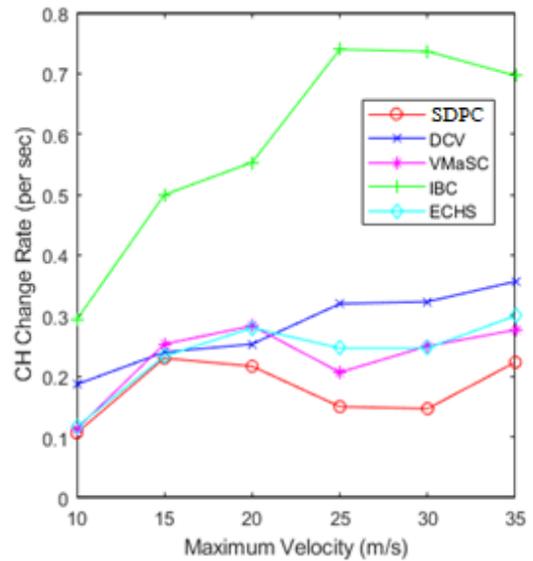

Fig. 6. Average CH change per second vs maximum vehicle velocity.

Fig. 6 shows the average CH change rate per second of SDPC compared to VMaSC, DCV, IBC and ECHS for different maximum velocity. The difference between MM and other algorithms varies for different maximum velocity; however, SDPC always shows lower CH change rate compared to other four algorithms. A lower CH change rate means the cluster is more stable. The difference between MM and other algorithms are nominal at 10 m/s and 15 m/s but significant at 20 m/s to 35 m/s. SDPC achieves 0.11 at 10 m/s to 0.2 at 35 m/s CH change rate per second where VMaSC and ECHS achieve 0.12 to 0.27, and DCV achieves 0.2 to 0.3 CH change rate per second for different maximum vehicle velocity.

Among VMaSC, ECHS and DCV; ECHS shows lower CH change rate when the maximum vehicle velocity is from 10 m/s to 20 m/s and VMaSC achieves lower CH change rate when the maximum vehicle velocity is from 25 m/s to 35 m/s.

The graph shows that only relative velocity or relative position or degree of node cannot always give the optimum stability for different maximum velocity. Rather combination of these parameters can provide optimized value.

Mohammad Mukhtaruzzaman

IBC achieves a range of 0.3 to 0.7 CH change rate. The curve for IBC shows much worse performance compared to SDPC, VMaSC, DCV, and ECHS, because of their forced cluster breakdown at the 2R distance from the intersection. In IBC, when a vehicle enters at the distance of twice of the transmission range from the intersection, vehicle breaks out from the existing cluster and form a new cluster. The number of new clusters can be one, two, or three since IBC creates three clusters within the 2R distance: one cluster for vehicles that will run straight direction after the intersection, one cluster for the vehicles that will run at the left direction after the intersection and the third cluster for the vehicles that will run at the right direction after the intersection. As a result, IBC can never provide an optimum CH change rate compared to other four algorithms.

*5.2  Average Duration of CH*

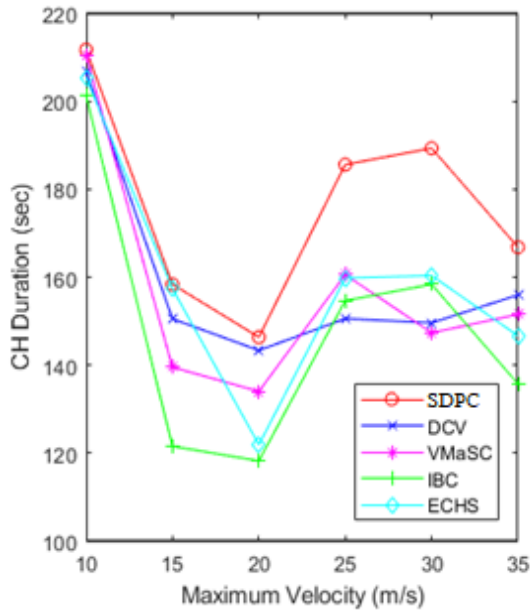

Fig. 7. Average CH duration vs maximum vehicle velocity.

Fig. 7 shows the average duration of CH for different maximum vehicle velocities. SDPC achieves a higher average CH duration compared to VMaSC, DCV, IBC and ECHS. MM achieves slightly better average CH duration when the maximum vehicle velocity is low from 10 m/s to 20 m/s and achieves much higher average CH duration when the maximum vehicle velocity is high from 25 m/s to 35 m/s. When the velocity of the vehicles are 25 m/s and 30 m/s, SDPC achieves 30, 32, 38 and 41 seconds higher average CH duration compared to ECHS, IBC, VMaSC and DCV respectively.

The graph shows that instead of fewer parameters if we use multiple parameters and consider movement at the intersection, we can achieve a higher CH average.

Moreover, we are selecting the CH from the direction-majority vehicles which contributed to increase the average duration of CH. From a vehicle cluster, some vehicles will run straight after the intersection, some vehicles will run at the left direction, and some vehicles will run at the right direction at the direction. The direction in which the most number of vehicles will be running after the intersection, we are calling this group of vehicles in the cluster as the majority-vehicles and choosing the CH from this group that helps SDPC to achieve a higher average CH duration compared to VMaSC, DCV, ECHS, and IBC.

*5.3  Average Duration of CM*

Fig. 8 shows the average duration of CM for different maximum vehicle velocities. From 10 m/s to 35 m/s, the average CM duration is continuously decreased because at the higher speed, the mobility of vehicles increases.

Except DCV, the performance of other four algorithms are merely in the same range while SDPC and IBC achieves slightly higher CM average than VMaSC and ECHS. Since DCV uses a smaller transmission range than the actual transmission range, its CH can cover a small number of vehicles resulting in a higher number of unattended vehicles which reflects in the average CM duration.

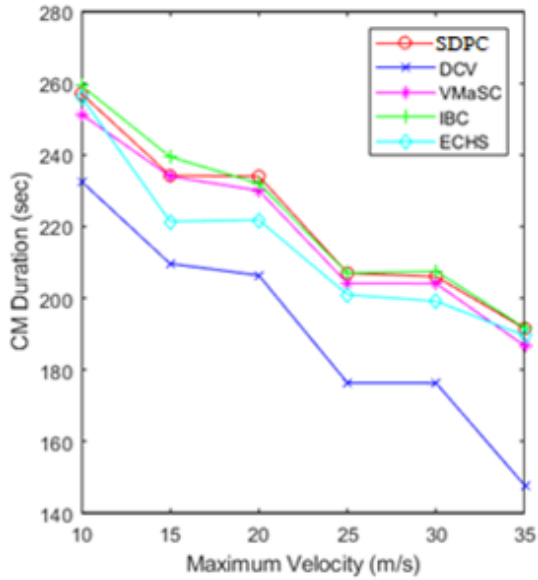

Fig. 8. Average CM duration vs maximum vehicle velocity.

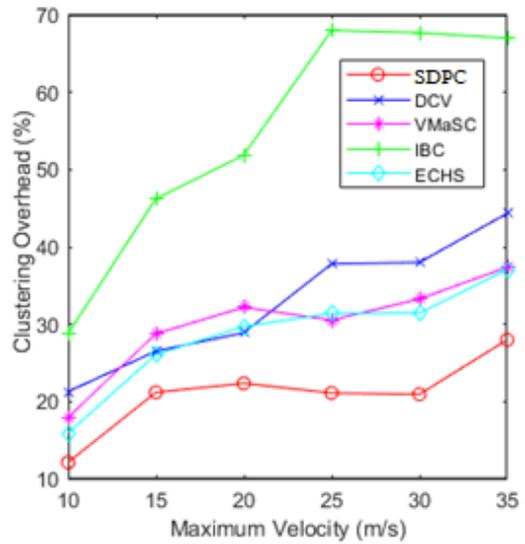

*5.4   Clustering Overhead*

Fig. 9. Clustering overhead vs maximum vehicle velocity.

Even though SDPC and IBC in this case is almost overlapping each other, IBC's performance for the average CM duration is not due to the strength of IBC algorithm, rather IBC forces vehicles to break from the existing cluster at a distance of 2R from the intersection. As a result, new vehicle is becoming available for vehicles to form a new cluster which was not able to join in a cluster due to out of the range of the CH.

Even though we do not consider the average CM duration as an important performance criterion for cluster stability, a slightly higher CM average achieved by MM compared to VMaSC, ECHS and DCV shows that while we improved the performance of CH change rate and the average CH duration, we did not compromise with the average CM duration also.

Fig. 9 shows the percentage of clustering overhead of SDPC, DCV, VMaSC, IBC and ECHS for different maximum vehicle velocities. From 10 m/s to 35 m/s, MM shows a significantly lesser number of clustering overhead compared to other four algorithms. MM achieves 11% at 10 m/s to 26% at 35 m/s. ECHS ranges from 16% to 35%, VMaSC ranges from 17% to 35% and DCV ranges from 21% to 44% at the speed of 10 m/s to 35 m/s. IBC rises from 30% at 10 m/s to 68% at 25 m/s, then remains nearly unchanged.

SDPC had achieved a lower number of CH change rate compared to VMaSC, DCV, ECHS, and IBC; at the same time, MM uses a lower number of states for the vehicles, and a lower number of state transitions for the vehicles. MM uses only three states for the vehicles and four state transitions. A lesser number of state transition and a lesser number of CH change contributed to a lower clustering overhead for SDPC.

Mohammad Mukhtaruzzaman

## 6 CONCLUSION AND FUTURE WORK

In this paper, we have introduced a novel stable dynamic feedback based VANET clustering algorithm. Instead of creating cluster based on the current road scenario, the future road scenario is constructed based on the continuous feedback from the vehicles and cluster if formed based on the constructed future road scenario. Also, neglecting the moving direction of the vehicle completely, an efficient approach of using the moving direction at the next intersection is considered in the cluster formation process. The relative position and movement of the vehicle at the next intersection are important in selecting the CH. When multiple vehicles' candidacy for the CH is equal, degree of vehicles is considered. Hence, the presented algorithm achieves higher stability, preventing frequent breaking of the clusters at the junction.

The presented algorithm shows superior performance in terms of average CH change rate, average CH duration, average CM duration, and clustering overhead. Generally, previous clustering algorithms made a trade-off between the performance at the straight road and at the intersection. On the contrary, the presented clustering algorithm achieves optimized performance at both straight road and intersection.

In the future, we want to implement our algorithm using a dozen of vehicles in real road to overcome any limitation to make it feasible for practical use.

Mohammad Mukhtaruzzaman